\documentclass[sn-mathphys-num]{sn-jnl}

\usepackage{graphicx}%
\usepackage{multirow}%
\usepackage{amsmath,amssymb,amsfonts}%
\usepackage{amsthm}%
\usepackage{mathrsfs}%
\usepackage[title]{appendix}%
\usepackage{xcolor}%
\usepackage{textcomp}%
\usepackage{manyfoot}%
\usepackage{booktabs}%
\usepackage{algorithm}%
\usepackage{algorithmicx}%
\usepackage{algpseudocode}%
\usepackage{listings}%

\theoremstyle{thmstyleone}%
\theoremstyle{thmstyletwo}%

\theoremstyle{thmstylethree}%

\begin{document}

\title[Article Title]{Can ChatGPT Overcome Behavioral Biases in the Financial Sector?
Classify-and-Rethink: Multi-Step Zero-Shot Reasoning in the Gold Investment}


\title[Article Title]{Can ChatGPT Overcome Behavioral Biases in the Financial Sector?
Classify-and-Rethink: Multi-Step Zero-Shot Reasoning in the Gold Investment}


\author*[1,2]{\fnm{Shuoling} \sur{Liu}}\email{sliudi@cse.ust.hk}

\author[3]{\fnm{Gaoguo} \sur{Jia}}\email{jiagaoguo@efunds.com.cn}
\equalcont{These authors contributed equally to this work.}

\author[3]{\fnm{Yuhang} \sur{Jiang}}\email{jiangyuhang@efunds.com.cn}
\equalcont{These authors contributed equally to this work.}

\author[3]{\fnm{Liyuan} \sur{Chen}}\email{chenly@efunds.com.cn}

\author[1]{\fnm{Qiang} \sur{Yang}}\email{qyang@cse.ust.hk}

\affil[1]{\orgname{Hong Kong University of Science and Technology}, \orgaddress{\city{Hong Kong}, \country{China}}}

\affil[2]{\orgname{CAS Key Laboratory of AI Safety, Institute of Computing Technology, Chinese Academy of Sciences}, \orgaddress{\city{Beijing}, \country{China}}}

\affil[3]{\orgdiv{Innovation Lab}, \orgname{E Fund Management Co., Ltd}, \orgaddress{\city{Guangzhou}, \country{China}}}

\abstract{Frame effect is a phenomenon observed in both humans and large language models where different descriptions of the same objectively identical problem can lead to different decisions. In behavioral finance, this effect must be carefully managed to prevent undesirable outcomes in the investment process. In the application of large language models, prompt engineering is often employed to mitigate the framing effect. While previous research in this domain has focused on sentiment classification or subject recognition, this study employs a large language model to score gold-related news and designs the "Classify-and-Rethink" prompt strategy from the perspective of behavioral finance. Our experiment demonstrates that the Classify-and-Rethink prompt approach effectively overcomes the framing effect and facilitates excess returns. Compared to alternative prompt designs, the Classify-and-Rethink strategy exhibits higher yields and Sharpe ratios. This framework provides a valuable basis for future research on behavioral finance in large-scale analysis of financial texts and offers investors a reliable means of mitigating behavioral biases.}




\maketitle

\section{Introduction}\label{sec1}

Large Language Models (LLMs) have achieved remarkable success recently, displaying exceptional capabilities in creating understandable and organized text ~\cite{wei2022emergent}. 
These LLMs have been utilized in diverse fields, such as clinical research, where domain-specific models like Med-Palm have achieved human-level performance~\cite{singhal2022large}. 
Recently, researchers have employed advanced prompt engineering to enhance the general reasoning ability of LLMs~\cite{Kojima2022LargeLM,Wei2022ChainOT}. 
Despite the remarkable success of zero-shot Chain-of-Thoughts (CoT) in solving general reasoning tasks, the potential of these methods still remains paid limited attention in the financial reasoning task.
To address this issue, we explore multiple prompt strategies and incorporated semantic news information to improve LLMs' performance on financial reasoning tasks.
To the best of our knowledge, we are the first to explore this important issue by applying ChatGPT to the gold investment.

In this work, our aim is to investigate the financial reasoning capabilities of LLMs and their capacity to generate logical and persuasive investment opinions. We will use ChatGPT, one of the most powerful LLMs recently, and prompt engineering to achieve this goal. Our research will focus on understanding the ability of LLMs in sophisticated analysis and reasoning within the context of investment decision-making. Our study finds that ChatGPT with CoT prompt can provide more explainable predictions and overcome behavioral biases, which is crucial in finance-related tasks and can achieve higher investment returns.

In conclusion, our main contributions are:
\begin{itemize}
    \item We propose a novel method named Classify-And-Rethink (CAR) to generate investment opinion using Chain-of-Thought Prompting. This method can enhance the financial reasoning ability of LLMs and be utilized in investment decisions.
    \item We evaluate our method by applying ChatGPT in the gold investment, and the back-testing results show that CAR outperforms the baseline both on the yield and Sharpe ratio.
    \item We conduct a comprehensive study of our CAR method, and find that our prompt method can help LLMs overcome behavioral biases in the financial sector.
\end{itemize}

\section{Method}\label{sec2}
\begin{figure*}
  \centering
  \includegraphics[width=0.9\linewidth]{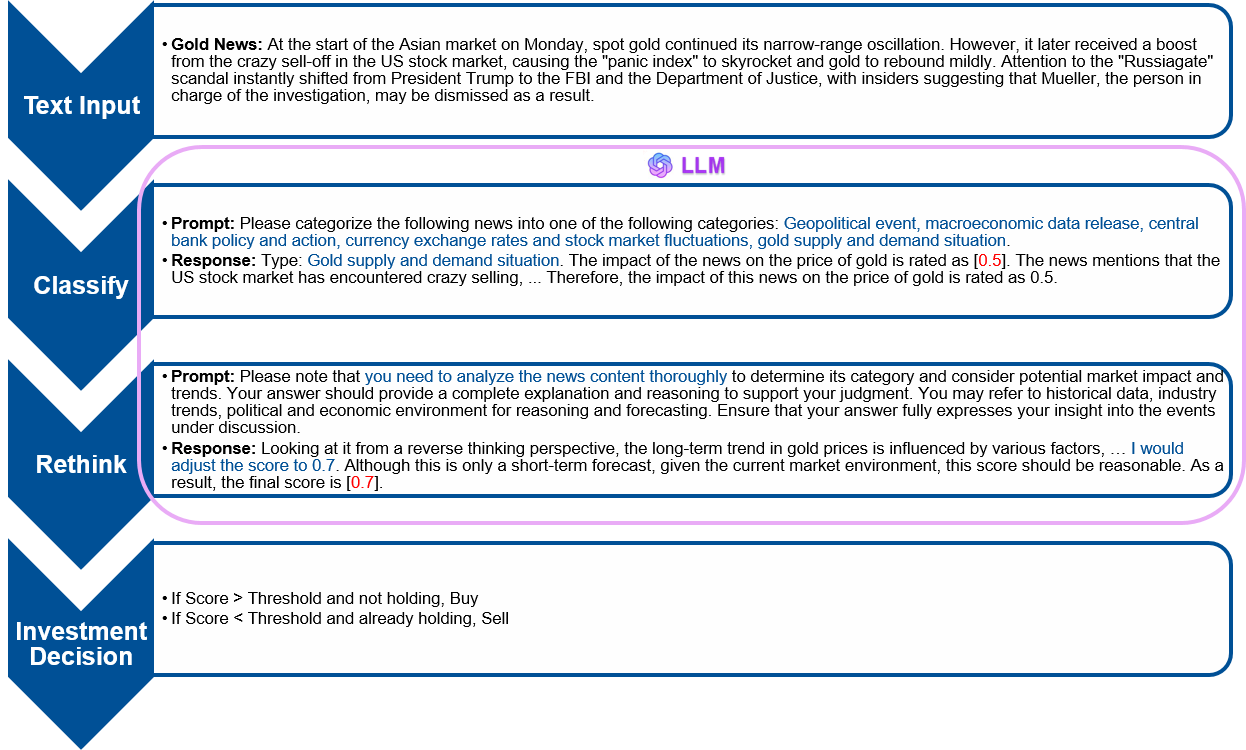}
  \caption{Framework of our method CAR. There are two main components: (1) Classify. The news is categorized into one of the six classes, then the LLM will give the opinion and the score of the news. (2) Rethink. The LLM will reflect on the reasonableness of the scoring from a long-term trend perspective, and make appropriate revisions to give the final score.}
  \label{fig:framework}
\end{figure*}

\subsection{Motivation} 
In the field of behavioral finance, investors' behavioral biases can be divided into two categories: cognitive errors and emotional biases. Compared to humans, large language models generally tend to maintain neutrality in their emotional responses. However, they are also susceptible to being misled by the way questions are asked. For humans, this is an example of the framing effect, and it is also an important reason why large models need to optimize their prompt words. "Framing effect" is a cognitive bias that refers to the influence of presentation format and method on decision-making. During the decision-making process, people often evaluate different options based on the format and method of the information presented, not just the inherent features of the decision itself. In finance, investors tend to focus too much on the short term or the information they have most access to, while ignoring the long-term and less sensational information. 

Here is a typical example of the framing effect in the financial field: when asked if a fund that suffered losses in a given year would rebound or continue to lose money the following year, more people thought it would rebound. Meanwhile, when the same group was asked to choose between two funds - one that made large profits for the first four years but suffered losses in the fifth, and another that made small profits in all five years - more people chose the second fund, as it was deemed more stable. This classic example of human behavioral bias can also influence large language models. It predicted that the fund in the first question was more likely to rebound while selecting the fund with profits in all five years for the second question.

Investors' behavioral biases are one of the sources of excess returns in quantitative finance. Listed companies may rely on the framing effect to decide how to disclose good and bad news to shareholders, and the news that investors receive often limits their investment decisions. As a highly scrutinized precious metal commodity, news related to gold can trigger faster price adjustments than most stocks. 
Therefore, relying solely on news sentiment to determine whether to buy may not keep up with market changes and may suffer from the influence of the framing effect, making it difficult to make the correct decision. As previously mentioned, the framing effect can also occur in large language models, so using models such as ChatGPT can simulate investors' behavioral biases and even correct them through the design of prompt word schemes.

\subsection{Datasets}
The relevant data used in this paper can be divided into two parts. 
The gold price used in the study is the daily closing price of the Shanghai Gold Exchange spot gold index(Au9999.SGE), which was obtained from the Wind database. 
The gold news was collected from http://www.dyhjw.com/ and covers Chinese gold news from 2018 onwards. Traditional sentiment recognition models generally rate such news as neutral because they usually only state objective facts and do not have obvious emotional features. 
However, these news items actually contain events related to supply and demand, central bank policy, and other factors closely related to gold prices. Using large models to process this type of news is appropriate because even in a zero-shot scenario, large models can still provide relatively universal economic laws. 
It should be noted that the impact of some events is very short-term and may be priced by the market within a single trading day. Without considering trades on a minute-by-minute basis, these short-term news items may influence investor decisions through framing effects. 
In the gold trading market, zero-shot large models can simulate ordinary investors affected by framing effects. Based on this, excess returns can be achieved by designing prompt word schemes.

\subsection{Prompt Design}
To convert a large amount of news into tradeable data, we use ChatGPT to score each news article and then process these scores. The scoring principle is based on the idea that news stories that are more favorable for future gold prices should receive higher scores, while those that are less favorable should receive lower scores, ranging from -1 to 1, and classified into levels of 0.1 points. The simplest approach is to write this requirement as an appropriate prompt and then score it.

In principle, using the very basic prompt word scheme described above, large language models will simulate the response of a relatively uninformed investor who receives daily news updates. Therefore, their actions are likely to be influenced by framing effects, and going against the framing effect may be more correct. However, most investors in the market have some financial knowledge and can make more detailed analyses of news items. Even in the case of zero-shot learning, large language models exhibit high classification accuracy. Based on this, we can fully utilize these existing models for automated news classification and scoring, to improve the efficiency and accuracy of news analysis. The news is categorized into six classes for scoring: geopolitical event, macroeconomic data release, central bank policy and action, currency exchange rates and stock market fluctuations, and gold supply and demand situation. Through our research, we have found that large language models can classify news articles with high accuracy. Furthermore, the explanations and scores generated by the model for each category are linguistically logical and economically reasonable, which differs from previous scoring methods.

Considering that the categorized model relies mainly on the logic of news for its thinking process, we propose using prompts to enable ChatGPT to think beyond the framework of the original news. Specifically, we aim to use prompts to make ChatGPT rethink on whether its previous scoring is reasonable from a long-term trend perspective. This approach allows the large model to move beyond simply repeating market sentiment and attempt to generate a score that is distinct from the market.

With the analysis above, we obtain the complete workflow of the CAR prompt framework as shown in Figure~\ref{fig:framework}: the first step involves news classification, the second step involves scoring based on the category, and the third step involves reflecting on the reasonableness of the scoring from a long-term trend perspective, and making appropriate revisions.

\subsection{Back-testing Strategy}
In order to present the experimental results clearly, we employed the most basic back-testing strategy. This strategy has only two states for holding gold, either an empty position or a full position. Starting with an empty position, we only buy gold and switch to a full position when the calculated indicators meet the buy conditions during a particular trading day when the position is empty. When the position is full, we only sell gold and switch to an empty position when the calculated indicators meet the sell conditions during a particular trading day.

The core of this back-testing strategy lies in how to calculate the buy and sell conditions. Large models are prone to behavioral biases and, without reflecting on them, their scores may be affected by framing effects similar to those experienced by ordinary investors. To obtain excess returns from this behavioral bias, we need to take the opposite approach. Using the score of every news article as a buy or sell criterion would have significant volatility, leading to an unstable trading strategy and high transaction fees. Therefore, we use the average score of the latest five news articles as the buy and sell standard. If the average score of the latest five news articles is below the threshold and the position is empty, we buy in with a full position; if the average score is above the threshold and the position is full, we sell out with an empty position.

When the large model reflects on this strategy from a long-term perspective, we can make adjustments from two aspects. Because we are computing for a longer-term impact, we take the average score of the latest 20 news articles. At the same time, we believe that at this point, the large model can use reflection to eliminate or even leverage framing effects, and the strategy will operate based on its scoring. If the average score of the latest 20 news articles is above the threshold and the position is empty, we buy in with a full position; if the average score is below the threshold and the position is full, we sell out with an empty position.

\section{Experiments}

\subsection{Details}
The experiment is dependent on the application programming interface (API) offered by OpenAI's GPT-3.5-turbo. For each news article, a unique dialogue is initiated. The Temperature parameter of the API is set to 0, the maximum length is set to 1024, and the presence penalty is set to 1 to guarantee consistent results, whereas all other parameters retain their default specifications. News reported after the Shanghai Gold Exchange's daily closure at 15:30 is attributed to the next trading day. The back-testing procedures start from January 2, 2018, and conclude on June 13, 2023.

\subsection{Main Results}
\begin{table}
    \centering
    \caption{Return and Sharpe Ratio of different strategies}
    \label{tab:result}
    \begin{tabular}{lrrr}
        \toprule
        Strategy  & Return & Sharpe Ratio  \\
        \midrule
        Buy-and-Hold     & 63.53\%    &0.811\\
        One-Step    & 63.44\%     &0.902\\
        Classify  & 73.41\%      &1.019\\
        Classify+Rethink (CAR)  & \textbf{80.35}\%     &\textbf{1.071}\\
        \bottomrule
    \end{tabular}
\footnotetext{Our CAR strategy outperforms other strategies both in returns and Sharpe ratio. As "One-Step" and "Buy-and-Hold" strategies show similar performance, we also achieve better results with only the "Classify" strategy.}
\end{table}

\begin{figure}
  \centering
  \includegraphics[width=0.9\linewidth]{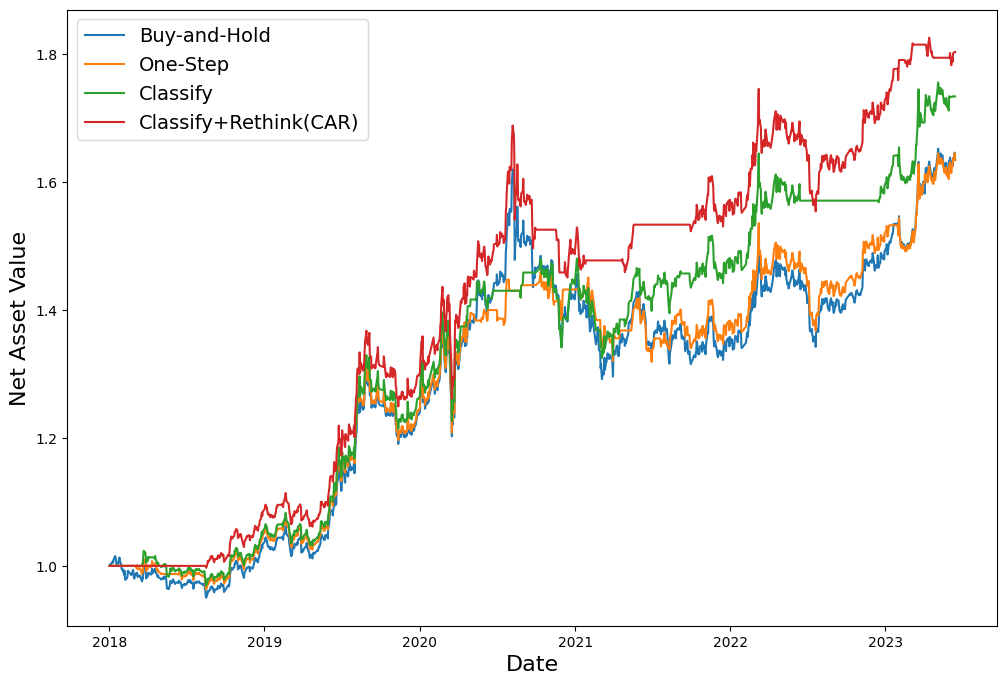}
  \caption{Asset value of different trading strategies without considering transaction costs. Our CAR method achieves the highest revenue as it can avoid significant drops as well as keep up with the upward trend when the gold price rises rapidly}
  \label{fig:result_fig}
\end{figure}
The experimental findings are outlined in Table~\ref{tab:result}, where three different strategies are contrasted. The "Buy-and-Hold" strategy involves continuously purchasing and holding gold from the onset of the back-testing period. On the other hand, the "One-Step" strategy assesses the average sentiment score based on the previous five news items utilizing the one-step prompt. If the score exceeds the predetermined value and holding, the gold is sold. Conversely, if it falls below the value and not holding, the gold is bought. Similarly, the "Classify" strategy ascertains the average sentiment score from the preceding five news items via the Two Steps prompt strategy. If the score surpasses the specified value and already holding, the gold is sold. By contrast, if it falls short of the value and not holding, the gold is bought. Finally, the "Classify-and-Rethink" strategy estimates the average sentiment score considering the latest 20 news items using the Rethinking prompt strategy. If the score exceeds the set value and not holding, the gold is purchased. In contrast, if it falls short of the value and already holding, the gold is sold. 

As depicted in Table~\ref{tab:result}, the CAR strategy yields superior returns and Sharpe ratio followed by the "Classify" strategy in second place. However, the "One-Step" and "Buy-and-Hold" strategies exhibit inconsequential performance differences.

Figure~\ref{fig:result_fig} presents the net value curves for the three strategies. As depicted in the figure, the "Classify" strategy averts the high gold position and curtails noteworthy drops. However, it misses the rapid market surge. Comparatively, the CAR strategy outperforms as it can keep pace with abrupt gold price hikes.

\begin{figure}
  \centering
  \includegraphics[width=0.9\linewidth]{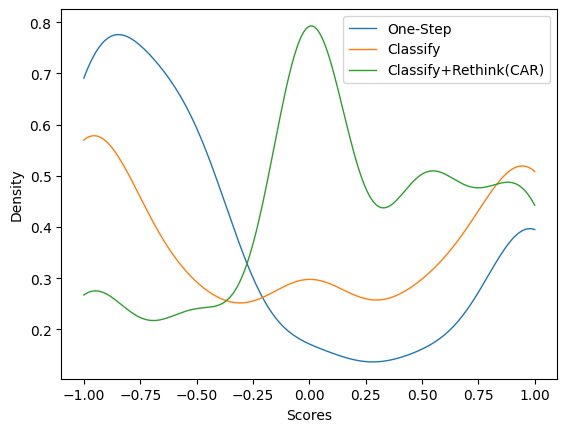}
  \caption{The distribution curve of scores given by different prompt strategies. The distribution of scores given by our CAR strategy is similar to the normal distribution, which is more aligned with the financial reality compared to the other two strategies.}
  \label{fig:result_cur}
\end{figure}

To further verify the effectiveness of the CAR strategy on news scoring, a statistical analysis of news scores under each strategy was carried out. Figure~\ref{fig:result_cur} illustrates that under the One-Step strategy, the scores are predominantly concentrated on a strong negative scale. In contrast, the Classify strategy exhibits a score distribution that gravitates toward two extremes. Conversely, the scores attained by the CAR strategy are closely aligned with a normal distribution. A normal distribution of scores not only aligns more precisely with financial reality, but also simplifies subsequent data processing. The score distribution implies that the CAR strategy can enable the model's overall scores to be more rational.

\subsection{Ablation Study}
We attempt to remove the classification process and directly make ChatGPT rethink score the original results. However, we found that the resulting strategy's profit was 42\%, which is lower than the most basic Buy-and-Hold strategy. This indicates that the classification module is crucial. When performing buy and sell operations at the same frequency randomly, the fluctuation of the achieved profitability ranged between 20\% and 40\%. This suggests that the strategy is more effective than randomly guessing. Moreover, the strategy demonstrated clear effectiveness and superiority, as it achieved higher profitability than holding onto gold continuously, even after executing 30 sell operations during a period of persistent upward gold trends over the past few years.

We also endeavor to utilize some open-source models, such as ChatGLM. Although it can provide insights on news-gold price trend connection akin to ChatGPT, it cannot generate specific scores based on prompt terms. Therefore, utilizing ChatGLM directly in comparative experiments is a challenge.

\subsection{Case Study}
\begin{table}[h]
    \centering
    \caption{Example responses of ChatGPT in different prompt strategies.}
    \label{tab:case_table}
    \begin{tabular}{p{1.8cm} c p{8.9cm}}
        \toprule
        Strategy  & Score & Explanation\\
        \midrule
        One-Step     & -0.8    & \small On Monday, spot gold in the Asian market continued narrow-range volatility in early trading. It rebounded slightly due to a boost from the crazy sell-off in the US stock market, leading to a surge in the "panic index". However, the shift in focus on the "Russian Gate" incident may lead to a decrease in demand for gold in the market, which could have a negative impact on gold prices.\\
        Classify    & 0.5     & \small Type: Gold supply and demand situation. The impact of the news on the price of gold is rated as [0.5]. The news mentions that the US stock market has encountered crazy selling, leading to a surge in the "fear index". This may cause investors' concerns about market instability, thereby increasing demand for safe-haven assets such as gold. However, the news also mentions that gold has only rebounded moderately, indicating that the market's demand for gold has not increased significantly. Therefore, the impact of this news on the price of gold is rated as 0.5.\\
        Rethink   & 0.7     & \small Looking at it from a reverse thinking perspective, the long-term trend in gold prices is influenced by various factors, including geopolitical events, macroeconomic data releases, central bank policies and actions, currency exchange rates, and stock market fluctuations. Although these factors were not mentioned in this article, we cannot rule out their potential impact on the price of gold. Therefore, I believe that rating this news as 0.5 may be too optimistic and requires appropriate adjustment. Considering that the US stock market sell-off may trigger safe-haven sentiment and push up gold prices, I would adjust the score to 0.7. Although this is only a short-term forecast, given the current market environment, this score should be reasonable. As a result, the final score is [0.7].\\
        \bottomrule
    \end{tabular}
\footnotetext{The "Classify" strategy correctly establishes the relationship between gold prices and the US stock market sell-off, while the "Rethink" strategy help reconsider the previous opinion and avoid the framing effect.}
\end{table}

Unlike traditional deep learning methods, using LLM to score news allows us to see the explanations behind its scores. Therefore, we can examine whether the basis of its scores is reasonable. Let's take a news article in 2018 as an example: "On Monday, spot gold in the Asian market continued its narrow-range volatility in the early session, and later received a boost from the crazy sell-off in the US stock market. The surge in the "panic index" caused gold to rebound slightly. The focus of attention on the "Russian Gate" incident has instantly shifted from President Trump to the FBI and the Justice Department, and sources said that the "Russian Gate" incident leader Mueller is likely to be dismissed as a result of this."

From Table~\ref{tab:case_table}, it can be seen that the explanation behind the score given by the "One-Step" strategy is not reasonable. It suggests that the Trump-Russia scandal would shift investors' attention and reduce demand for gold, which is not necessarily true. The "Classify" strategy correctly establishes the relationship between gold prices and the US stock market sell-off as well as the surge in the "fear index", but its assertion that gold demand did not increase significantly is not well-supported by evidence. Although we expect the "Rethink" strategy to take a reverse thinking perspective, it still gives a more positive score and focuses on inducing safe-haven sentiment, which is likely to have a greater impact on future prices. We can see that CoT and rethinking did indeed help the large language model avoid the framing effect.

\section{Discussion}
We attempt to use prompt words in both Chinese and English, with the results being very similar, with score differences typically less than 0.1 points. However, we have yet to systematically compare the differences between news in different languages. From both language and economic perspectives, language is unlikely to have a significant impact on the method we proposed. Nevertheless, there may be significant differences in effectiveness when prompt words are relatively simple.

Furthermore, the stability of the strategy is a topic that deserves future discussion. Currently, our experiments are based solely on historical data. As large language models continue to evolve, it is worth conducting further research to observe whether strategies that were previously able to generate excess returns will regress to the mean over time, as assets become more fully priced. In such a market environment, it is crucial to identify ways to ensure market participants' interests, particularly for relatively weak individual investors, which is an ethical model that financial professionals applying large language models must face.

\section{Related Work}
\subsection{Large Language Models}
In recent years, the field of large language models (LLMs) has seen remarkable advancements, with these models demonstrating powerful performance in text completion and the ability to adapt to a wide range of NLP tasks after fine-tuning. To better align with user interaction habits, InstructGPT~\cite{ouyang2022training} employs reinforcement learning from human feedback (RLHF)~\cite{christiano2017deep} and Instruction Fine-Tuning (IFT)~\cite{wei2021finetuned} during the training process, which enhances the model’s performance and adaptability across various tasks and instructions.
In the financial domain, BloombergGPT~\cite{wu2023bloomberggpt} represents a significant milestone. This 50 billion parameter language model, trained on extensive financial data, significantly outperforms existing models on financial tasks without compromising its performance on general LLM benchmarks, signaling a major advancement in the integration of LLMs and finance.
ChatGPT\footnote{https://openai.com/blog/chatgpt}, a fine-tuned variant of InstructGPT, possesses extraordinary capabilities in language processing and responding to user requests. Given its impressive performance and accessibility, it is applied in our CAR strategy with the anticipation that it can effectively overcome behavioral biases.

\subsection{Financial Reasoning}
The financial industry has seen a rapid adoption of natural language processing (NLP) research, as highlighted in a study by ~\cite{wu2023bloomberggpt}. However, a vast majority of research in finance has primarily focused on token or sequence classification tasks, as demonstrated in works such as ~\cite{araci2019finbert,shah2022flue}. These studies have limited the exploration of reasoning abilities within the field, posing challenges for the development of specialized language models with finance-specific reasoning capabilities.

Some research has shown that the large language models with 6B parameters can exhibit logical and consistent financial reasoning abilities. Moreover, the effectiveness of these capabilities can be further enhanced through larger datasets and improved instruction tuning, as discussed in the research by ~\cite{son2023removing} and ~\cite{Son2023BeyondCF}.

In addition, ~\cite{xie2023wall} argues that while ChatGPT struggles to outperform traditional machine learning techniques in zero-shot multimodal stock price prediction tasks, there is potential for improvement in this area.
Overall, while NLP research has made significant contributions to the financial industry, there is still room for exploration and enhancement of the financial reasoning ability.

\subsection{Prompt Engineering}
The response of large language models (LLMs) is deeply influenced by the quality of the prompt design, with the Chain-of-Thought (CoT) approach being a pivotal technique that enhances this capability. Wei et al.~\cite{Wei2022ChainOT} introduced Chain-of-Thought prompting, which involves several reasoning steps before providing an answer to the input question, thereby significantly enhancing the reasoning proficiency of LLMs. This few-shot prompting technique has been further refined in subsequent research, including optimizing the prompt format~\cite{Chen2022ProgramOT}, selecting effective prompts~\cite{Lu2022DynamicPL}, ensembling different prompts~\cite{Weng2022LargeLM}, and decomposing complex problems~\cite{Press2022MeasuringAN}.
Chen et al.~\cite{Chen2022ProgramOT} proposed the PoT prompting, which utilizes LLMs with code pre-training to write a program as a rationale, separating computation from reasoning. Kojima et al.~\cite{Kojima2022LargeLM} introduced Zero-shot-CoT, which enables the generation of reasoning steps without the need for exemplars, while Auto-CoT~\cite{Zhang2022AutomaticCO} makes use of demonstration examples to minimize manual effort.
Zheng et al.~\cite{zheng2023progressive} introduced Progressive-Hint Prompting (PHP), a novel prompting method that facilitates automatic, multiple interactions between users and LLMs. This method employs previously generated answers as hints to iteratively guide the user towards the correct answer. This progressive approach to hint-based prompting represents an innovative way to enhance the interactive and reasoning capabilities of LLMs.


\section{Conclusion}
In this study, we aim to evaluate the effectiveness of ChatGPT as a tool for financial decisions. 
We propose a method named CAR to enhance the financial reasoning ability of LLMs by overcoming behavioral biases.
To evaluate our method, we apply CAR by using ChatGPT to generate investment opinions and scores based on the gold news, and then use it to make financial decisions. 
The back-testing experiments show that the use of the CAR strategy helps large models to overcome the framing effect and score news, and the scoring distribution of news is more reasonable compared to unprocessed large models, leading to higher investment returns.
From case analysis, the CAR strategy enables models to avoid being affected by some nearby but less important news information and to focus on the most critical factors that affect gold prices, thereby providing more reasonable scoring.
These findings suggest that ChatGPT can provide plausible and credible explanations for the impact on the gold price of gold news.

In conclusion, our study explores the potential of ChatGPT in the financial reasoning task and shows the spark for future research and application of LLMs in financial investment.

\section*{Declarations}
\subsection*{Funding}
Not applicable
\subsection*{Conflict of Interest}
The authors declare the following financial interests relationships which may be considered as potential conflicts of interest:
Shuoling Liu, Gaoguo Jia, Yuhang Jiang and Liyuan Chen are both affiliated with E Fund Management Co., Ltd. Their affiliation with E Fund Management Co., Ltd presents a financial interest given the nature of the research.
No other potential conflicts of interest relevant to this article are reported.
\subsection*{Ethics Approval and Consent to Participate}
Not applicable
\subsection*{Data Availability}
The gold news data is provided by the financial data service company Datayes, and they have not given their permission for researchers to share their data. Data requests can be made to Datayes via https://mall.datayes.com/datapreview/4015. The gold price of Au9999 data is provided by the financial data service company Wind, and they have not given their permission for researchers to share their data. Data requests can be made to Datayes via DFSupport@wind.com.cn. Other data generated during the current study are available from the corresponding author on reasonable request. 
\subsection*{Materials Availability}
Materials generated during the current study are available from the corresponding author on reasonable request.
\subsection*{Code Availability}
Code generated during the current study are available from the corresponding author on reasonable request.

\bibliography{sn-article}

\end{document}